\documentclass{an}
\usepackage{graphicx}
\usepackage{txfonts}
\usepackage{gensymb}
\usepackage{verbatim}
\usepackage{natbib}
\usepackage[switch]{lineno}

\newcommand{\SUB}{}
\begin{document}
\hyphenation{ab-rupt-ly}
\title{Time series analysis of long--term
photometry of BM Canum Venaticorum
\thanks{The analysed photometry and numerical results of the analysis
are both published electronically at the CDS via anonymous ftp to
cdsarc.u-strasbg.fr (130.79.128.5) or via
http://cdsarc.u-strasbg.fr/viz-bin/qcat?J/A+A/yyy/Axxx}}
\author{
L. Siltala           \inst{1}
\and L. Jetsu        \inst{1}
\and T. Hackman      \inst{1}
\and G.W. Henry      \inst{2}
\and L. Immonen      \inst{1}
\and P. Kajatkari    \inst{1}
\and J. Lankinen     \inst{1}
\and J. Lehtinen     \inst{1}
\and S. Monira       \inst{1}
\and S. Nikbakhsh    \inst{1}
\and A. Viitanen     \inst{1}
\and J. Viuho        \inst{1}
\and T. Willamo      \inst{1}
}
\institute{Department of Physics,
P.O. Box 64, FI-00014 University of Helsinki, Finland
\and
Center of Excellence in Information Systems, Tennessee State University,
3500 John A. Merritt Blvd., Box 9501, Nashville, TN 37209, USA}
\date{Received / Accepted}
  \abstract
   {
%
    We study standard Johnson differential $V$ 
    photometry
    of the RS CVn binary \object{BM CVn}
    spanning over a quarter of a century.
    Our main aims are
    to determine the activity cycles,
    the rate of surface differential rotation and the rotation period
    of the active longitudes
    of \object{BM CVn}.
    The Continuous Period Search (CPS) is applied to the photometry.
    The changes of the mean
    and amplitude of the light curves are used to search for activity cycles.
    The rotation period changes give an
    estimate of the rate of surface differential rotation.
    The Kuiper method is applied to the epochs of the primary and
    secondary minima
    to search for active longitudes.
    The photometry reveals the presence of a stable mean light curve (MLC)
   connected to the orbital
    period $P_{\mathrm{orb}}=20\fd6252$ of this binary.
    We remove this MLC from the original $V$ magnitudes which gives us
    the corrected $V'$ magnitudes.
    These two samples of $V$ and $V'$ data are analysed separately with CPS.
    The fraction of unreliable CPS models decreases when the MLC is removed.
    The same significant activity cycle of approximately 12.5 years
    is detected in both $V$ and $V'$ samples.
    The estimate for the surface differential rotation coefficient,
    $k \ge 0.10$, is the same for
    both samples, but the number of unrealistic period estimates
    decreases after removing the MLC.
    The same active longitude period of
    $P_{\mathrm{al}}=20\fd511 \pm 0\fd005$
    is detected in the $V$ and $V'$ magnitudes.
    This long--term regularity in the epochs of primary
    and secondary minima of the light curves
    is not caused by the MLC. On the contrary,
    the MLC hampers the detection of active longitudes.
    }

   \maketitle
%


\section{Introduction}

\cite{Bid83} noticed the strong \ion{Ca}{ii} K line emission of
\object{BM CVn} (\object{HD116204}, \object{BD+39 2635}).
\cite{Hal83} suspected photometric variability.
This was confirmed by \cite{Boy84b},
who discovered a period of $P_{\mathrm{phot}}=21\fd3$
from a light curve having an amplitude of $0.\!\!^{\mathrm m}07$.
{Furthermore, \cite{Boy84b} suggested that BM CVn is a RS CVn binary.}
Among other published $P_{\mathrm{phot}}$ values are
$21\fd9$ \citep{Moh87},
$20\fd66\pm0\fd03$ \citep{Str89}
and
$20\fd2\pm0\fd5$ \citep{Erd09}.
However, \citet{Koe02} detected
no periodicity in the Hipparcos
photometry of $n=169$ observations.

\cite{Gri88} showed that \object{BM CVn} is a binary with
$P_{\mathrm{orb}}=20\fd6252\pm0\fd0018$,
and noted that the secondary companion could not be detected
spectroscopically.
\object{BM CVn} is currently classified as a single-lined eclipsing binary
with $v \sin{i}=15$ km/s in the Third catalogue of
chromospherically active binaries \citep{Eke08}.
The suggested spectral types of the primary are
K3~III \citep{Sim87},
KI~III \citep{Sat90}
and
G8~III \citep{Bof93}.
\citet{Moh87} arrived at a lower limit estimate,
$R\ge 6 R_{\odot}$, for the primary.
\citet{Sta94} argued that the inclination of the orbital
plane, $i_{\mathrm{orb}}=26\degree$, is nearly equal to
the inclination of the rotation axis of the primary,
$i_{\mathrm{rot}}=24\degree$.
Their estimate for the primary radius was $15 \pm 2 R_{\odot}$.

Activity induced chromospheric or coronal emission of
\object{BM CVn} has been detected in the optical
\ion{Ca}{ii} H\&K and H$\alpha$ lines,
as well as in UV and X-ray wavelengths
\citep{Fek86,Str90,Sat90,Dem93b,Fra94,Mon00,Per11}.
Radio and IR emission has also been detected
\citep{Mit96,Hel99,Haa09}.

{Recent studies have shown that for some RS CVn binaries the light curve is 
heavily biased by the ellipsoidal shape of the primary 
\citep{Roe15,Roe15b}. Furthermore, other physical processes, e.g. mass 
transfer between close binaries could distort the light curve. These
kinds of distortions would always follow the periodicity of the orbital period.
Long-term active longitudes, on the other hand, may follow a different 
periodicity \citep{Hac11}. Thus, if the difference between these periods
is large enough, we can separate the spot activity from other variability
connected to the orbital period. One way to attempt this is to subtract a
mean light curve fit calculated with the orbital period. This approach
is, in a sense, similar to the pre-whitening used in detecting stellar
differential rotation \citep{Rei13}.}

In the current study, we apply the Continuous Period Search
\citep[][CPS]{Leh11} to photometry of \object{BM CVn}
spanning over a quarter of a century. 
{One of our aims is to look for active longitudes. Therefore we
check whether the binary has any orbital period mean light curve, 
which could distort
the detection of active longitudes.}

   \begin{figure}
   \centering
   \includegraphics[width=\hsize]{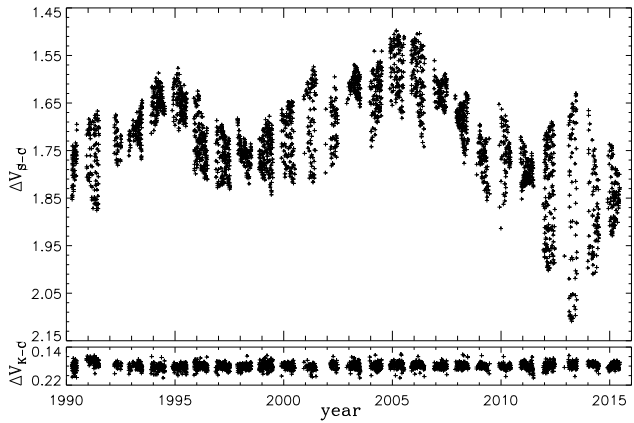}
      \caption{Photometric data.
               Upper panel: All $\Delta V_{\mathrm{S-C}}$ data.
               Lower panel: All $\Delta V_{\mathrm{K-C}}$ data in the
               same magnitude scale as in the upper panel}
         \label{Data}
   \end{figure}

\section{Observations}

Our differential photometry was obtained with the T3 0.4m automated
photoelectric telescope (APT) at Fairborn Observatory in Arizona.
The observations were made between
April 6th, 1990 (HJD = 2447987.9)
and
June 15th, 2015 (HJD = 2457188.8).
The comparison and check stars of our target star, S = \object{BM CVn},
were
C = \object{HD~116010}  \citep[][K2II-III, V=5.60]{Yos97}
and
K = \object{HD~115271}  \citep[][A7V, $V=5.78$]{Roy07}.
The mean ($m$) and the standard deviation ($s$) of
standard Johnson differential magnitudes,
$\Delta V_{\mathrm{S-C}}$ (Fig. \ref{Data}, 
upper panel: $n=2930$),
were $m \pm s= 1.\!\!^{\mathrm m}721 \pm    0.\!\!^{\mathrm m}096$.
The respective values for the $\Delta V_{\mathrm{K-C}}$
differential magnitudes (Fig. \ref{Data}, 
lower panel: $n=2744$)
were  $m \pm s= 0.\!\!^{\mathrm m}1797 \pm 0.\!\!^{\mathrm m}0064$.
Hence, C = \object{HD~116010} was a reliable comparison star,
the accuracy of our photometry was approximately $\sigma_{\mathrm{V}}=0.\!\!^{\mathrm m}0064$,
and the observed variations of S = \object{BM CVn} were certainly real.
The photometric data reduction procedures and the
operation of the T3 0.4~m APT have been 
described in detail, e.g., by \cite{Hen99} and \cite{Fek05}.
The differential $\Delta V_{\mathrm{S-C}}$ magnitudes are hereafter referred
to as the original $V$ magnitudes.

\section{CPS method \label{cps}}

We analysed the original $V$ magnitudes of \object{BM CVn} with the CPS method
formulated by \citet{Leh11}.
This method is described here only briefly,
because we have already applied it to the photometry
of numerous stars \citep{Leh11,Hac11,Hac13,Leh12,Kaj14,Kaj15,Leh16}.
We divided the observations into datasets having a maximum
length of $\Delta T_{\mathrm{max}} = 1.5 P_{\mathrm{orb}} = 30\fd94$.
The modelled datasets contained
at least $n \ge n_{\mathrm{min}}=14$ observations.
CPS uses a sliding window with a length of $\Delta T_{\mathrm{max}}$
and models all datasets having at least $n_{\mathrm{min}}$ observations
within such a window.
The notation for the mean of the $n$ observing times $t_i$
of a dataset is $\tau$.
The CPS model is
\begin{eqnarray}
\hat{y}(t_i) =
\hat{y}(t_i,\bar{\beta}) =
M + \sum_{k=1}^K{[B_k\cos{(k2\pi ft_i)} + C_k\sin{(k2\pi ft_i)}]},
\nonumber
\label{model}
\end{eqnarray}
where $K$ is the model order
and $\bar{\beta}=[M,B_1,..,B_K,C_1,...,C_K,f]$ is vector of free parameters.
The best modelling order $K$ for each dataset is chosen
by using a Bayesian criterion \citep[][Eq. 6]{Leh11}.
Here, the tested orders for this best model were ${0 \le K \le 2}$.
These models gave the following physically meaningful
light curve parameters:
the period $P(\tau)$,
the mean brightness $M(\tau)$,
the peak to peak amplitude $A(\tau)$,
and
the epochs of the primary and secondary minima in time,
$t_{\mathrm{min,1}}(\tau)$ and $t_{\mathrm{min,2}}(\tau)$.
The error estimates for these light curve parameters were determined
with the bootstrap method \citep{Leh11}.
The reliable and unreliable parameter estimates
were identified from these bootstrap results.
The parameter estimates were considered reliable
if the distribution of the bootstrap estimates
for all model parameters and the model residuals
followed a Gaussian distribution \citep{Leh11}.
Two concrete examples of the use of this reliability criterion
can be found in \cite[][Figs. 2 and 4]{Jet96c}.
We use the notation $R(\tau)=0$ for the reliable light curve parameter estimates
and $R(\tau)=1$ for the unreliable ones.
The modelling results correlate for temporally overlapping subsets
containing common data.
We selected a sequence of independent datasets which did not overlap.
The notation for these independent datasets is ${\rm IND}(\tau)=1$,
while that for the remaining overlapping datasets with common data is ${\rm IND}(\tau)=0$.

   \begin{figure*}
   \centering
   \includegraphics[width=\hsize]{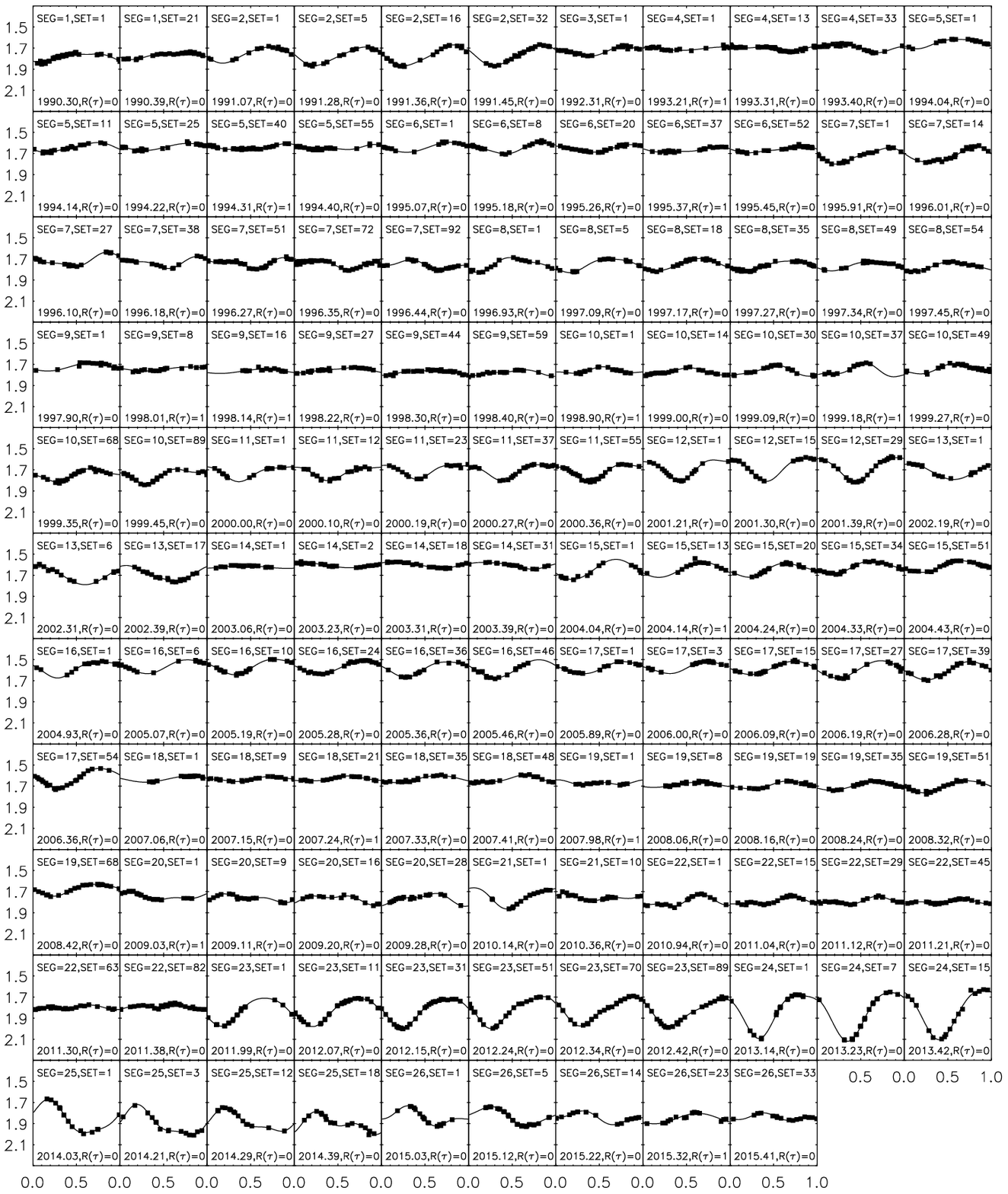}
      \caption{The data and the light curves of 119 independent 
$V$ magnitude datasets (${\mathrm{IND}}(\tau=1)$).
               The reliable and unreliable models a
re denoted with ${\mathrm{R}}(\tau)=0$ and 1, respectively.
The x-axis is the phase $\phi$ and the y-axis is the 
magnitude $V$.
               The computation of the phases $\phi$
is explained in the last paragraph
               of Sect. \ref{origcpsresults}.
}
         \label{Lightcurves}
   \end{figure*}

\section{Analysis of $V$ magnitudes}

\subsection{CPS results for the original $V$
 magnitudes \label{origcpsresults}}

We detected periodicity in all 1319 original $V$ magnitude datasets.
The order of the best model was
$K=1$ in 205 datasets and $K=2$ in 1114 datasets.
CPS gave the following numbers of different types of
$M(\tau)$, $A(\tau)$, $P(\tau)$, $t_{\mathrm{min,1}}(\tau)$
and $t_{\mathrm{min,2}}(\tau)$ estimates:

\begin{center}
\addtolength{\tabcolsep}{-0.12cm}
\begin{tabular}{cllll}
\hline
                          & ${\rm IND}(\tau)\!=\!1$ & ${\rm IND}(\tau)\!=\!1$ &
                            ${\rm IND}(\tau)\!=\!0$ & ${\rm IND}(\tau)\!=\!0$ \\
                          & ${\rm R}(\tau)\!=\!0$ & ${\rm R}(\tau)\!=\!1$ &
                            ${\rm R}(\tau)\!=\!0$ & ${\rm R}(\tau)\!=\!1$ \\
\hline
                   $M(\tau)$ &  $n=107$ {$[\blacksquare]$}  &   $n=12$ {$[\Box]$}    &  $n=1098$ {\scriptsize $[\times]$} &  $n=102$ {\scriptsize $[\times]$} \\
                   $A(\tau)$ &  $n=107$ {$[\blacksquare]$}  &   $n=12$ {$[\Box]$}    &  $n=1098$ {\scriptsize $[\times]$} &  $n=102$ {\scriptsize $[\times]$} \\
                   $P(\tau)$ &  $n=107$ {$[\blacksquare]$}  &   $n=12$ {$[\Box]$}    &  $n=1098$ {\scriptsize $[\times]$} &  $n=102$ {\scriptsize $[\times]$} \\
  $t_{\mathrm{min,1}}(\tau)$    &  $n=107$ {$[\blacksquare]$}   &  $n=12$ {$[\Box]$}     &  $n=1098$ {\scriptsize $[\times]$} &  $n=102$ {\scriptsize $[\times]$} \\
  $t_{\mathrm{min,2}}(\tau)$    &  $n=23$  {$[\blacktriangle]$} &   $n=6$ {$[\triangle]$}&   $n=293$ {\scriptsize $[\times]$} &  $n=42$ {\scriptsize $[\times]$} \\
\hline
\end{tabular}
\addtolength{\tabcolsep}{+0.12cm}
\end{center}
\noindent
Note that the symbols given in 
the brackets are those used in Figs. \ref{MeAm}--\ref{Orbi} 
for the modelling results of the original 
$V$ magnitude data. {\SUB All CPS analysis results for the independent 
$V$ datasets are published electronically at the CDS.}

The data and the CPS models of
the 119 independent datasets are shown in Fig. \ref{Lightcurves}.
We first computed the phases 
$\phi_1={\rm FRAC}[(t-t_{\mathrm{min,1}}(\tau))/P(\tau)]$,
where  ${\rm FRAC}[x]$ removes the integer part of its argument $x$.
Then, the phases $\phi_{al,1}$ of the primary minima $t_{\mathrm{min,1}}(\tau)$
were computed with the active longitude ephemeris of Eq. \ref{AcLo}.
The photometry and the CPS models are plotted
as a function of the phase $\phi=\phi_1+\phi_{al,1}$ in Fig. \ref{Lightcurves}.

\subsection{Activity cycles in $V$ photometry \label{Cycle}}

\citet{Hor86} formulated the power spectrum method (hereafter PSM)
which was applied by \citet{Bal95} to search for activity cycles
in the chromospheric Ca~II H\&K emission line data.
For example,
\citet{Rod00} and \citet{Kaj15}
have applied PSM to the following light curve parameters:
$M(\tau)$                   (axisymmetric part of spot distribution),
$A(\tau)$                   (non-axisymmetric part of spot distribution),
$M(\tau) + A(\tau)/2$         (maximum spottedness) or
$M(\tau) - A(\tau)/2$         (minimum spottedness).
These four parameters are also analysed here,
although there are actually only two independent variables,
$M(\tau)$ and $A(\tau)$.
Furthermore, the amplitude of the light curve does not depend
only on the area of the spots, but there are also other causes,
e.g. variations of the temperature of the spots 
or the connection between their projected area and their latitude.

\begin{figure}
   \centering
   \includegraphics[width=\hsize]{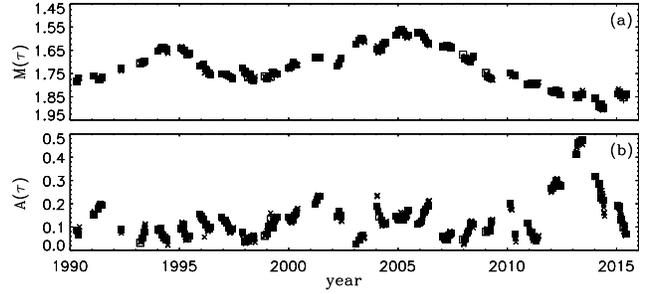}
      \caption{{\bf a}) Mean $M(\tau)$ for the original $V$ magnitudes,
               {\bf b}) Amplitude $A(\tau)$ for the original $V$ magnitudes.
               The symbols are explained in the first
               paragraph of Sect. \ref{origcpsresults}
              }
         \label{MeAm}
   \end{figure}

\begin{figure}
   \centering
   \includegraphics[width=\hsize]{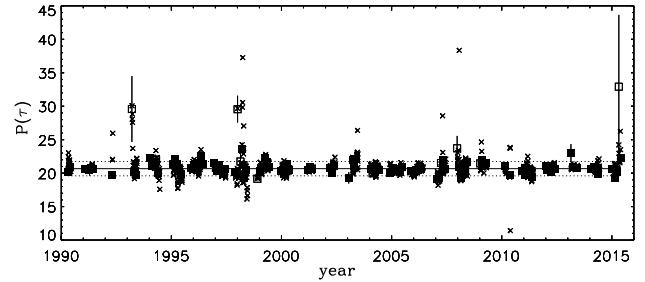}
      \caption{Period $P(\tau)$ for the original $V$ magnitudes.
The $P_{\mathrm{w}}$ and $P_{\mathrm{w}} \pm 3 \Delta P_{\mathrm{w}}$ 
levels are denoted
with horizontal continuous and dotted lines, respectively.}
         \label{Pe}
   \end{figure}

The $M(\tau)$ and  $A(\tau)$ changes of \object{BM CVn} are shown
in Figs. \ref{MeAm}.
PSM detected the following activity cycles $P_{\mathrm{c}}$ in the $n=107$
independent and reliable estimates of the above
four parameters of \object{BM CVn},
where $F$ is the false alarm probability \citep{Hor86}
\begin{center}
\begin{tabular}{ccc}
                    & $P_{\mathrm{c}} \pm \sigma_{P_C}$   & $F$                  \\
$M(\tau)$           & $13.\!\!^{\mathrm y}1 \pm 0.\!\!^{\mathrm y}2$  & $5 \times 10^{-11}$ \\
$A(\tau)$           & $12.\!\!^{\mathrm y}0 \pm 0.\!\!^{\mathrm y}3$  & $2 \times 10^{-4}$ \\
$M(\tau)+A(\tau)/2$ & $12.\!\!^{\mathrm y}8 \pm 0.\!\!^{\mathrm y}2$  & $5 \times 10^{-10}$ \\
$M(\tau)-A(\tau)/2$ & $18.\!\!^{\mathrm y}2 \pm 0.\!\!^{\mathrm y}4$  & $7 \times 10^{-9}$ \\
\end{tabular}
\end{center}
\noindent
All $F$ values indicate that these cycles are very significant.
Considering the $\sigma_{P_C}$ errors, the first three parameters
may follow the same activity cycle of approximately $12.\!\!^{\mathrm y}5$ which
has been repeated twice between 1990 and 2015.
However, the parameter $M(\tau)-A(\tau)/2$ appears to follow another
cycle of $18.\!\!^{\mathrm y}2$.

\subsection{Differential rotation in $V$ 
photometry \label{Diff}}

The $P(\tau)$ changes of \object{BM CVn} are shown in Fig. \ref{Pe}.
Numerous unrealistic period estimates were
obtained for low amplitude light curves,
e.g. in the years 1992, 1993, 1995 and 1998.
Our estimate for the differential rotation of \object{BM CVn} is
based on the $n=107$ reliable $P(\tau)$ estimates of independent 
datasets. We compute the value of the parameter
\begin{eqnarray}
Z=6 \Delta P_{\mathrm{w}}/P_{\mathrm{w}}
=0.10 \equiv 10\%,
\label{zestimate}
\end{eqnarray}
where
$P_{\mathrm{i}} \pm \sigma_{\mathrm{P,i}}$ is period of $i$:th dataset, 
$w_{\mathrm{i}}=\sigma_{\mathrm{P,i}}^{-2}$ is the weight,
$P_{\mathrm{w}}=[\sum_{\mathrm{i=1}}^{107} w_{\mathrm{i}} P_{\mathrm{i}}]/
[\sum_{\mathrm{i=1}}^{107}w_{\mathrm{i}}]$ is the weighted mean of $P_{\mathrm{i}}$
and
$\Delta P_{\mathrm{w}}=
\{[\sum_{\mathrm{i=1}}^{107} w_{\mathrm{i}}
(P_{\mathrm{i}}-P_{\mathrm{w}})^2]/[\sum_{\mathrm{i=1}}^{107}w_{\mathrm{i}}]\}^{1/2}$
is the error of this weighted mean $P_{\mathrm{w}}$ 
\citep[][Eq. 14]{Leh11}.
The numerical values for \object{BM CVn} are 
$P_{\mathrm{w}} \pm \Delta P_{\mathrm{w}}
=  20\fd67 \pm 0\fd36$.
We use the multiplying constant value of 6 in Eq. \ref{zestimate},
because parameter $Z$ is the $\pm 3 \Delta P_{\mathrm{w}}$
upper limit for the period variations.

\citet[][their Eq. 15]{Leh11} introduced the relation
$Z_{\mathrm{phys}}=Z^2 -Z^2_{\mathrm{spu}}$,
where $Z$ measures the observed period changes of Eq. \ref{zestimate},
while $Z_{\mathrm{phys}}$  and $Z_{\mathrm{spu}}$
measure the real physical and unreal spurious period changes.
The mean of the half amplitude  $A(\tau)/2$ of
the above mentioned 107 models of \object{BM CVn}
is $A_{\mathrm{half}}=0.\!\!^{\mathrm m}068$.
The data precision is $N=\sigma_{\mathrm{V}}=0.\!\!^{\mathrm m}0064$.
This yields a
``signal to noise'' ratio of $\epsilon =A_{\mathrm{half}}/N \approx 10$.
Together with the mean number of data points per dataset, 
$n_{\rm data}=19.9$, and the ratio of the mean rotation period 
to the dataset length, $n_{\rm rot}=1.45$, this predicts spurious changes of 
$Z_{\rm spu} \approx 0.08$ \citep[][Eq. 10]{Leh16}. 
This is a major fraction of the raw estimate $Z=0.10$, 
although the physically originating component of the period 
changes may still be estimated 
at $Z_{\mathrm{phys}} \approx (Z^2-Z_{\mathrm{spu}}^2)^{1/2} = 0.06$.

One approximation for the solar law of differential
rotation is $P(b)=P(b=0)/[1-k_{\odot}(\sin{b})^2]$,
where $b$ is the latitude
and $k_{\odot}=0.2$ is the solar differential rotation coefficient.
If this law were valid for \object{BM CVn} and
its $P(\tau)$ were reliable tracers of surface differential rotation,
the differential rotation coefficient
of \object{BM CVn} would be $|k| = Z_{\mathrm{phys}}/h$,
where $b_{\mathrm{min}}$ and  $b_{\mathrm{max}}$ are
the minimum and maximum latitudes of spot activity, and
$h=\sin^2{b_{\mathrm{max}}}-\sin^2{b_{\mathrm{min}}}$
\citep{Jet00}.
The exact latitudes of the spots can not be determined
from photometric observations, 
and thus the numerical 
value of $h$ remains unknown.
For example, if spots form at all latitudes between the equator
and pole of \object{BM CVn},
this coefficient reaches its maximum value $h=1$.
Thus, the relation $|k| > Z_{\mathrm{phys}} \approx 0.06$ is valid
for all possible $b_{\mathrm{min}}$ and  $b_{\mathrm{max}}$ values.

\subsection{Active longitudes in $V$ 
photometry \label{Active} }

The spots on the surface of rapidly rotating giants have been observed to
concentrate on long-lived active longitudes \citep[e.g.][]{Jet96a}.
Such structures have been detected, e.g.
with the nonweighted or weighted Kuiper 
test formulated in \citet[][Sect. 3.1]{Jet96b}.
In this test,
the phases $\phi_{\mathrm{i}}$ of $n$ 
time points $t_{\mathrm{i}}$ are first
computed with the tested period $P$.
These phases are then arranged into 
increasing order (i.e. rank order).
The monotonously increasing sample distribution function 
$F_{\mathrm{n}}(\phi_{\mathrm{i}})=i/n$ of these phases
is compared to
the sample distribution function of an even 
distribution $F(\phi)=\phi$, i.e. a random distribution.
The Kuiper test statistic is $V_{\mathrm{n}}=D^{+}+D^{-}$,
where 
$D^{+}=F_{\mathrm{n}}(\phi)-F(\phi)$ 
and
$D^{-}=F(\phi)-F_{\mathrm{n}}(\phi)$. 
A large $V_{\mathrm{n}}$ value indicates that 
the phases $\phi_{\mathrm{i}}$ do not represent 
a sample drawn from a random distribution,
i.e. the phases $\phi_{\mathrm{i}}$ are not evenly distributed and
there is periodicity in time points $t_{\mathrm{i}}$ 
with the tested period $P$.

We applied the nonweighted Kuiper
test to the reliable $t_{\mathrm{min,1}}(\tau)$ estimates
of all independent datasets ($n=107$).
The tested period range was between
$0.85 P_{\mathrm{W}}=17\fd6$ and
$1.15 P_{\mathrm{W}}=23\fd8$.
The best rotation period
for the active longitudes of \object{BM CVn} was
$P_{\mathrm{al}}=20\fd511 \pm 0\fd005$.
This periodicity reached an extreme significance
of $Q_{\mathrm{K}}=1 \times 10^{-7}$
\citep[][Eq. 24]{Jet96b}.
We also applied the same test to the reliable
$t_{\mathrm{min,1}}(\tau)$ and  $t_{\mathrm{min,2}}(\tau)$
estimates of all independent datasets ($n=130$).
The result was exactly the same,
$20\fd511 \pm 0\fd008$ ($Q_{\mathrm{K}}=4 \times 10^{-7}$).

\begin{figure}
   \centering
   \includegraphics[width=\hsize]{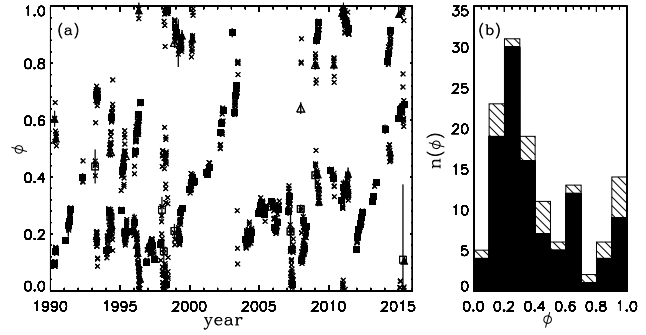}
      \caption{{\bf (a)} $t_{\mathrm{min,1}}(\tau)$ and  $t_{\mathrm{min,2}}(\tau)$ 
for the original $V$ magnitudes
               with the active longitude ephemeris of Eq. \ref{AcLo}.
               {\bf (b)} Histogram of the number of
               values $n(\phi)$ within 0.1 bins in phase. 
               The dark and shaded
               areas denote the reliable
$t_{\mathrm{min,1}}(\tau)$ and  $t_{\mathrm{min,2}}(\tau)$ estimates
               of independent datasets, respectively.
              }
         \label{MiMi}
   \end{figure}

All $t_{\mathrm{min,1}}(\tau)$ and
 $t_{\mathrm{min,2}}(\tau)$ estimates of \object{BM CVn} are
shown in Fig. \ref{MiMi}.
The phases are calculated from
the active longitude ephemeris
\begin{eqnarray}
\mathrm{HJD}_0 = 2~447~987.8762+20\fd511{\mathrm{E}},
\label{AcLo}
\end{eqnarray}
\noindent
where the zero epoch is the time of the first photometric observation.
The majority of the reliable
primary minima of independent datasets concentrate between phases
0.1 and 0.6 (Fig. \ref{MiMi}a: closed squares and
Fig. \ref{MiMi}b: dark areas).
The corresponding simultaneous secondary minima
are usually about half a rotation
apart from these primary minima
(Fig. \ref{MiMi}a: closed triangles).
The light curves had only one minimum
in the years 1991--1992, 1997, 2001--2006, 2011--2013, i.e.
two minima were present for half of the time.
When the activity was present on both longitudes,
the longitude with stronger activity determined the
phase of $t_{\mathrm{min,1}}(\tau)$ (Fig. \ref{MiMi}a: closed squares).
Sometimes the activity shifted abruptly
nearly to the opposite side of the stellar hemisphere,
like in the years 2003--2004.
This type of the ``flip-flop'' events have previously been
observed, e.g. on the rapidly rotating single G4~III
giant FK Comae \citep{Jet93}.

   If the inclination of the primary were indeed only $i_{\mathrm{rot}}=24 \degr$ \citep{Sta94},
then extreme spot coverage would be required to explain
the high amplitudes of $A(\tau) \approx 0.\!\!^{\mathrm m}5$ in the year 2013 (Fig. \ref{MeAm}).
%

Using the Barnes-Evans relation as formulated by \cite{Lacy1977} we can 
derive an 
alternative estimate of the stellar radius with the formula
\begin{eqnarray}
\log {R/R_\odot} = 7.4724 - 0.2 V_0 - 2 F_V + \log d
\end{eqnarray}
\noindent
where $F_V = 3.977 - 0.429 (V-R)_0$ and $[d] = pc$. The Hipparcos parallax of 
8.86 mas \citep{vanleeuwen2007} combined
with the $V$ magnitude and
$V-R \approx 0.81$ \citep{Eke08} indicate a primary radius of  
$\sim 8 R_{\odot}$.
This would mean that that the rotational inclination of 
the primary is in fact $i_{\mathrm rot} \sim 50 \degr$.
This value would
still require a high, but not extreme spot coverage. Furthermore, assuming 
approximately the same orbital inclination, would result in a secondary
mass of $\sim 0.3 M_{\odot}$. Thus it is no surprise that its spectral lines 
have not been observed.

 Recently,
\citet{Roe15} showed that the ellipsoidal shape of
the primary of another RS CVn binary \object{$\sigma$ Gem}
\citep[][$P_{\mathrm{d}}\!=\!19\fd604471 \pm 0\fd000022$]{Dum97}
offered an alternative explanation for the active longitude hypothesis
presented by \citet{Kaj14}.
For \object{BM CVn},
the active longitude period
$P_{\mathrm{al}}=20\fd511\pm0\fd005$ is only 0.55\% smaller
than
the orbital period
$P_{\mathrm{orb}}=20\fd6252\pm0\fd0018$.
This period difference causes a $\Delta \phi =2.48$ phase difference during the whole time span of data.
The three periods of \object{BM CVn} increase in the order
$P_{\mathrm{al}} <P_{\mathrm{orb}} < P_{\mathrm{w}}$.
The $P_{\mathrm{w}}=20\fd67$ error,
$\Delta P_{\mathrm{w}}=0\fd36$,
does not exclude the cases $P_{\mathrm{w}}<P_{\mathrm{orb}}$ or $P_{\mathrm{w}} < P_{\mathrm{al}}$.
However,
the accuracy of the other two periods excludes the case $P_{\mathrm{al}} > P_{\mathrm{orb}}$.

We decided to test the ellipsoidal primary shape hypothesis for \object{BM CVn}.
The orbital period ephemeris in
\citet{Gri88} was
\begin{eqnarray}
\mathrm{HJD}_0=2~445~252.12+20\fd6252{\mathrm{E.}}
\label{Elli}
\end{eqnarray}

   \begin{figure}
   \centering
   \includegraphics[width=\hsize]{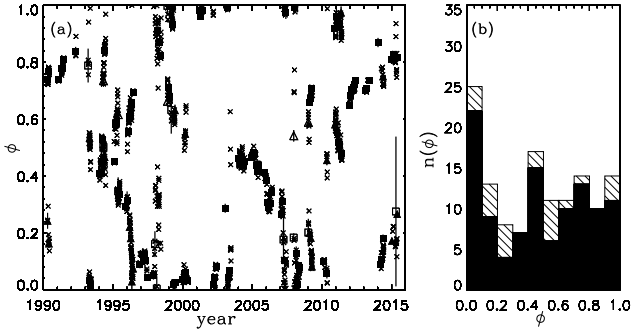}
      \caption{$t_{\mathrm{min,1}}(\tau)$ and  $t_{\mathrm{min,2}}(\tau)$ 
for the original $V$ magnitudes
               with the orbital period ephemeris of Eq. \ref{Elli}. 
Otherwise as in Fig. \ref{MiMi}.
              }
         \label{Orbi}
   \end{figure}

   \begin{figure}
   \centering
   \includegraphics[width=\hsize]{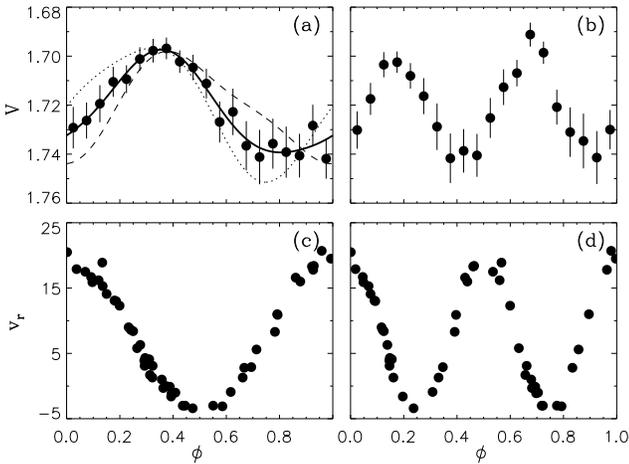}
   \caption{{\bf (a)} The MLC of \object{BM CVn}
               with the ephemeris Eq. \ref{Elli}.
               The bin width was 0.050 in phase $\phi$.
               The error bars show the mean error within each bin.
               The continuous line shows the best fit with the model 
           of Eqs. \ref{modelmlc}.
               {The dashed and the dotted lines show the MLC
               for the binned data before and after the mid epoch
               2002.85.}
               {\bf (b)} The MLC for the period $2P_{\mathrm{orb}}$ in Eq. \ref{Elli}.
               {\bf (c)} Radial 
                 velocities $v_{\mathrm{r}}$ in \citet{Gri88}
                ~with the ephemeris Eq. \ref{Elli}.
               {\bf (d)} Same data for the period $2P_{\mathrm{orb}}$ in Eq. \ref{Elli}.
              }
         \label{BinVr}
    \end{figure}

\noindent
The $t_{\mathrm{min,1}}(\tau)$ and  $t_{\mathrm{min,2}}(\tau)$ 
phases of \object{BM CVn}
computed with this ephemeris are shown in Fig. \ref{Orbi}.
These phases show no clear concentrations at $\phi=0.25$ or 0.75,
which could be one signature of ellipticity.
The minima of the binned $B$ and $V$ magnitude light curves of
\object{$\sigma$ Gem}
were at these particular orbital phases in \citet[][their Fig. 5]{Roe15}.
The results of a similar binning of the $V$ magnitudes
of \object{BM CVn} are shown in Fig. \ref{BinVr}a.
This  $0.\!\!^{\mathrm m}042$ peak to peak amplitude
mean light curve (hereafter MLC) shows only one minimum and one maximum,
while there were two maxima and minima in the MLC of
\object{$\sigma$ Gem}.

The MLC of \object{BM CVn} would also have two minima and maxima,
if the real $P_{\mathrm{orb}}$ of \object{BM CVn} were two times larger
than the one reported in \citet{Gri88} (Fig. \ref{BinVr}b).
However, this double period hypothesis must be rejected,
because the radial velocity measurements from \citet{Gri88}
follow a single and double sine wave
with $P_{\mathrm{orb}}$ and $2P_{\mathrm{orb}}$,
respectively (Figs. \ref{BinVr}cd).
These results do not support the ellipsoidal shape hypothesis,
because the MLC of \object{BM CVn} does not have two minima and maxima.
We will later discuss the possible cause of this sinusoidal MLC
of \object{BM CVn} in the end of Sect. \ref{Conc}.
Whatever the real cause may be,
it is certain that this regular $0.\!\!^{\mathrm m}042$ variation
has been misleading our CPS analysis of the original $V$ magnitudes.

{We also computed the MLC for the first and second part of the data,
i.e. before and after the mid epoch of 2002.85.
These two MLCs are denoted with dashed and dotted lines in Fig. \ref{BinVr}a.
Both of these curves show one maximum and minimum.
The maximum deviation of both curves from the MLC of all
data is only $0.\!\!^{\mathrm m}014$.
The phase and the height of the MLC maximum of \object{BM CVn}
has remained very stable,
but the depth and the phase of the MLC minimum has varied.
These MLC minimum variations are at least partly caused by the high amplitude
light curves of the years 2012 and 2013.
However, the MLC of \object{BM CVn} in Fig. \ref{BinVr}a (continuous line)
must be a real phenomenon,
because the erratic changes of the light curve
mean, amplitude and period,
and especially those of the minimum and maximum phases,
should cause a constant long--term MLC.
The MLC was computed with $P_{\mathrm{orb}}$.
This induces a phase difference of $\Delta \phi =2.48$
with $P_{\mathrm{al}}$ during the whole time span of data.
Hence, the observed
MLC phase coherence is not caused by active longitudes.
}

\subsection{MLC corrected $V'$ magnitudes}

We used the Bayesian criterion from \citet[][Eq. 6]{Leh11}
to determine the best modelling order $K$ for
the binned $V$ magnitudes of Fig. \ref{BinVr}a.
The tested orders were $0 \le K \le 4$.
The best order was $K=2$.
Hence, we modelled these binned $V$ magnitudes with
\begin{eqnarray}
g_2(\phi,\bar{\beta}_{\mathrm{MLC}}) 
=  ~a_0 & \!+\!& a_1\cos{(2\pi \phi)}
   +b_1\sin{(2\pi \phi)} \label{modelmlc} \\
        & \!+\! & a_2\cos{(4\pi \phi)}+b_2\sin{(4\pi \phi)}. \nonumber
\end{eqnarray}
where the free parameters were
 $\bar{\beta}_{\mathrm{MLC}}=[a_0,a_1,b_1,a_2,b_2]$ and
the orbital phases $\phi$ were calculated 
from the ephemeris of Eq. \ref{Elli}.
The best fit had
$a_0 = 1.721 \pm 0.002$,
$a_1 = 0.011 \pm 0.002$,
$b_1 =-0.017 \pm 0.003$,
$a_2 = 0.000 \pm 0.002$
and
$b_2 = 0.003 \pm 0.002$.
The continuous line outlines this model in Fig. \ref{BinVr}a.
The Bayesian criterion from \citet[][Eq. 6]{Leh11}
revealed that a second order component was present in the MLC,
although the amplitude of this second order part was low,
i.e. constant $a_2$ was zero and constant $b_2$ was very close to zero.
We used this same criterion to determine the best
$K$ value for the CPS models of {\it all} 
$V$ and $V'$ datasets.

This best fit of Eq. \ref{modelmlc} was used to
remove the MLC from the $V$ magnitudes of \object{BM CVn}.
The corrected magnitudes were computed from
\begin{eqnarray}
V'(t_i) & = & V(t_i) - g_2(\phi,\bar{\beta}_{\mathrm{MLC}}) + a_0
\label{procedure}
\end{eqnarray}
where $V(t_i)$ were the original data
and the values of the free parameters
$\bar{\beta}_{\mathrm{MLC}}$ were those of this best fit 
of Eq. \ref{modelmlc}.
The CPS analysis results do not depend on subtracting a constant
value from  differential photometry.
Therefore, we did not subtract the MLC mean, i.e. the constant
$a_0 =1.721$, from the $V(t_i)$ data.
In other words, $a_0$ was ``added back'' in Eq. \ref{procedure}.

\begin{figure}
   \centering
   \includegraphics[width=\hsize]{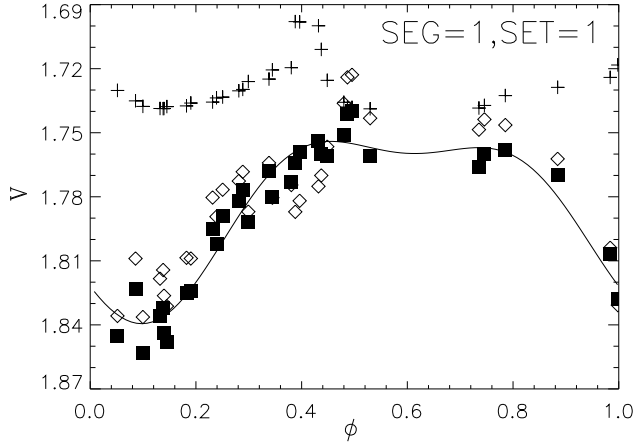}
      \caption{An example of the MLC correction procedure.
               The data (closed squares) and the CPS model (continuous line)
               are from Fig. \ref{Lightcurves} (SEG=1, SET=1).
               The crosses denote MLC magnitudes computed 
               from Eq. \ref{modelmlc}.
               The diamonds show the MLC corrected $V'$ magnitudes
               (Eq. \ref{procedure}).
              }
         \label{Subtracted}
   \end{figure}

One example of this
correction procedure of Eqs. \ref{modelmlc} and \ref{procedure}
is displayed in Fig. \ref{Subtracted}.
The original $V$ data are denoted with closed squares (SEG = 1, SET = 1).
The continuous line shows the light curve for these original data,
i.e. the first light curve from Fig. \ref{Lightcurves}.
The MLC values $g_2(\phi,\bar{\beta}_{\mathrm{MLC}})$ computed
with Eq. \ref{modelmlc} are scattered 
(Fig. \ref{Subtracted}: crosses).
The reason for this scatter is that these MLC values
are computed with the period $P_{\mathrm{orb}}=20\fd6252$,
but
the
correct period for these data is much 
shorter
$P(\tau)=20\fd14\pm0\fd31$.
This mixes the MLC phases when 
these phases are computed ``incorrectly'' with $P(\tau)$,
i.e. not with the correct period $P_{\mathrm{orb}}$.
The upward peak in the MLC values (crosses) 
close to phase 0.35 provides a nice example 
of the correction procedure.
When this peak is subtracted from the original $V$ data (closed squares),
the corrected $V'$ data show a dip at the same phase (open diamonds).

  \begin{figure*}
   \centering
   \includegraphics[width=\hsize]{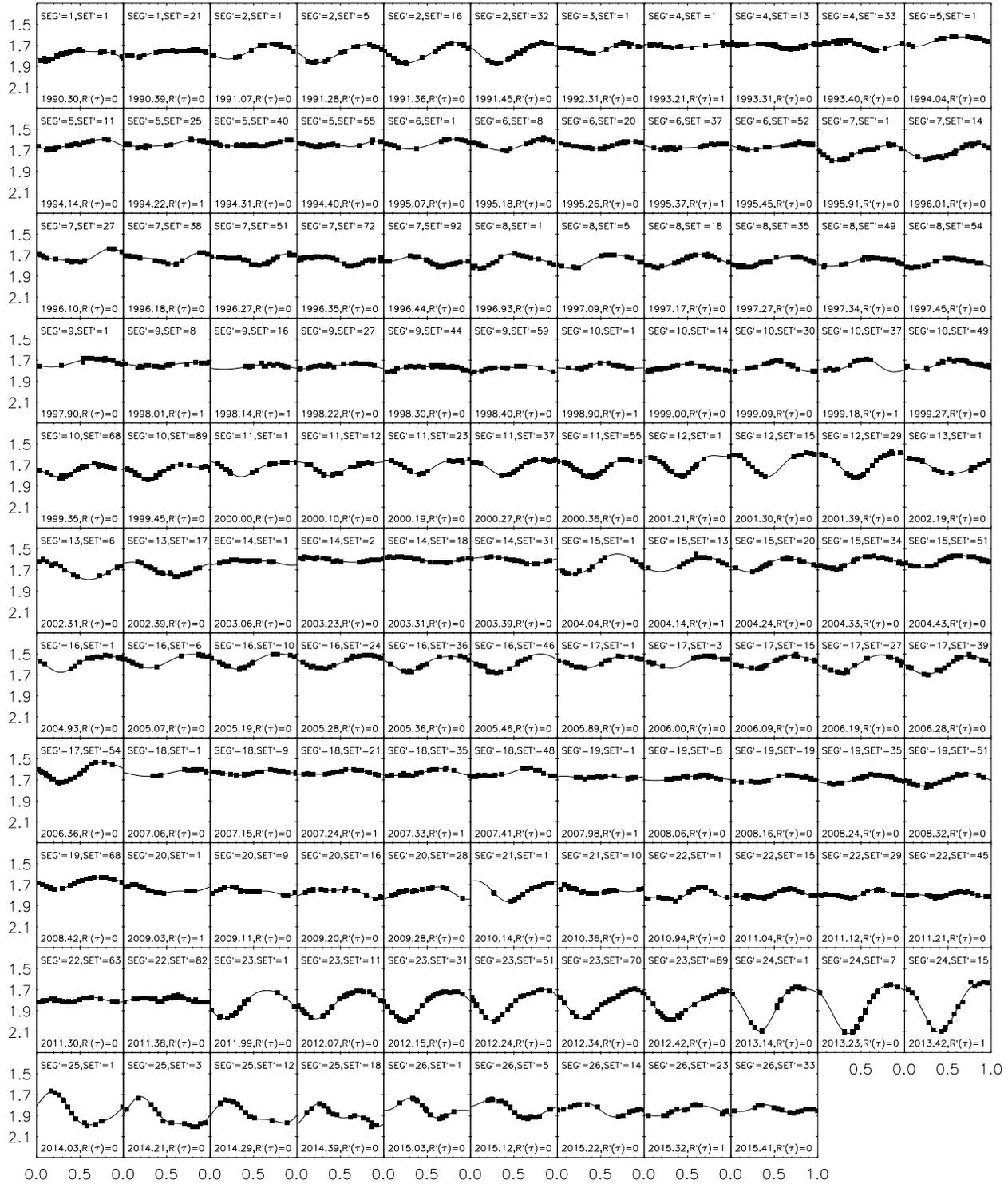}
      \caption{The light curves of 119 independent $V'$ magnitude
               datasets (${\rm IND'}(\tau=1)$).
               The reliable and unreliable models are denoted with ${\mathrm{R'}}(\tau)=0$ and 1, respectively. 
The x-axis is the phase $\phi$ and the y-axis is the 
magnitude $V'$. The computation of phases $\phi$
is explained the last paragraph of Sect \ref{convcpsresults}.}
         \label{Lightcurvesconv}
   \end{figure*}

\section{Analysis of $V'$ magnitudes}

\subsection{CPS results for the MLC corrected 
$V'$ magnitudes \label{convcpsresults}}

\begin{figure}
   \centering
   \includegraphics[width=\hsize]{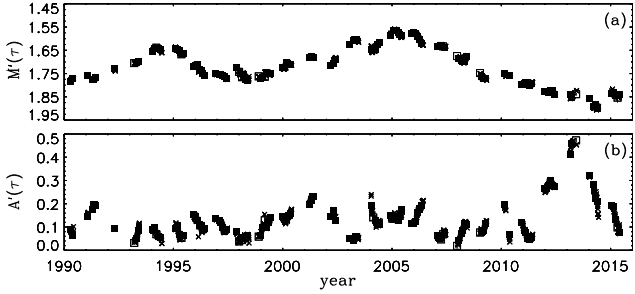}
      \caption{{\bf a}) Mean $M'(\tau)$ for the $V'$ magnitudes.
               {\bf b}) Amplitude $A'(\tau)$ for the $V'$ magnitudes.
               The symbols are explained in the first paragraph
               of Sect. \ref{convcpsresults}
              }
         \label{MeAmconv}
   \end{figure}

We use a prime ($'$) to denote all CPS analysis results
for the $V'$ magnitudes, e.g. $M'(\tau)$ for the mean.
For $\tau$, this notation is not required,
because the observing times $t_i$
of the
${\mathrm{V}}(t_i)$
and
${\mathrm{V'}}(t_i)$
magnitudes are the same.
The total number of CPS models for the corrected $V'$ magnitudes was 1319.
Again, CPS detected periodicity in all datasets.
The order of the best model was
$K'=1$ in 220 datasets and $K'=2$ in 1099 datasets.
The number of different types of CPS parameter estimates,
$M'(\tau)$, $A'(\tau)$, $P'(\tau)$, $t'_{\mathrm{min,1}}(\tau)$ and
$t'_{\mathrm{min,2}}(\tau)$, were:

\begin{center}
\addtolength{\tabcolsep}{-0.12cm}
\begin{tabular}{cllll}
\hline
                       & ${\rm IND'}(\tau)\!=\!1$ & ${\rm IND'}(\tau)\!=\!1$ &
                         ${\rm IND'}(\tau)\!=\!0$ & ${\rm IND'}(\tau)\!=\!0$ \\
                       & ${\rm R'}(\tau)\!=\!0$ & ${\rm R'}(\tau)\!=\!1$ &
                         ${\rm R'}(\tau)\!=\!0$ & ${\rm R'}(\tau)\!=\!1$ \\
\hline
                   $M'(\tau)$ &  $n'=105$ {$[\blacksquare]$}  &   $n'=14$ {$[\Box]$}    &  $n'=1117$ {\scriptsize $[\times]$} &  $n'=83$ {\scriptsize $[\times]$} \\
                   $A'(\tau)$ &  $n'=105$ {$[\blacksquare]$}  &   $n'=14$ {$[\Box]$}    &  $n'=1117$ {\scriptsize $[\times]$} &  $n'=83$ {\scriptsize $[\times]$} \\
                   $P'(\tau)$ &  $n'=105$ {$[\blacksquare]$}  &   $n'=14$ {$[\Box]$}    &  $n'=1117$ {\scriptsize $[\times]$} &  $n'=83$ {\scriptsize $[\times]$} \\
  $t'_{\mathrm{min,1}}(\tau)$    &  $n'=105$ {$[\blacksquare]$}  &    $n'=14$ {$[\Box]$}     &  $n'=1117$ {\scriptsize $[\times]$} &  $n'=83$ {\scriptsize $[\times]$} \\
  $t'_{\mathrm{min,2}}(\tau)$    &  $n'=24$ {$[\blacktriangle]$} &    $n'=5$ {$[\triangle]$}&   $n'=321$ {\scriptsize $[\times]$} &  $n'=33$ {\scriptsize $[\times]$} \\
\hline
\end{tabular}
\addtolength{\tabcolsep}{+0.12cm}
\end{center}
\noindent
The symbols used for
the modelling results of these parameters
in Figs.  \ref{MeAmconv}--\ref{MiMiconv}
are given above in 
the brackets.
The fraction of unreliable CPS models was 114/1319 = 8.6\% for the
original $V$ magnitudes.
This fraction decreased to 97/1319 = 7.4\% for the $V'$ magnitudes.
Hence, the number
of reliable models (${\rm R'}(\tau)=0$)
increased when the MLC was removed from
the original $V$ magnitudes.
{\SUB The CPS analysis results of the corrected $V'$ magnitudes
are also published only electronically at the CDS.}

The CPS models of 119 independent $V'$ magnitude data\-sets
are shown in Fig. \ref{Lightcurvesconv}.
The phases were first computed from
$\phi'_1={\rm FRAC}[(t-t'_{\mathrm{min,1}}(\tau))/P'(\tau)]$.
Then, we computed the phases $\phi'_{al,1}$ of the
primary minima $t'_{\mathrm{min,1}}(\tau)$
with the active longitude ephemeris of Eq. \ref{AcLo}.
The $V'$ magnitudes and  the CPS models are plotted
as a function of the phase $\phi'=\phi'_1+\phi'_{al,1}$
in Fig. \ref{Lightcurvesconv}.
The light curves of the original $V$ and corrected $V'$ magnitudes
are nearly identical (Figs. \ref{Lightcurves} and \ref{Lightcurvesconv}).
The largest change in the mean values $M(\tau)$ and $M'(\tau)$
is $0.^{\mathrm{m}}014$. It occurs in SET=18 of SEG 11.
The largest amplitude change, $0.^{\mathrm{m}}027$, 
between $A(\tau)$ and $A'(\tau)$
occurs in SET=1, SEG=19.
The light curves of this latter dataset are displayed
in Figs. \ref{Lightcurves} and \ref{Lightcurvesconv}.

\subsection{Activity cycles in $V'$ photometry \label{Cycleconv}}

The amplitude of the MLC was low, i.e. only $0.\!\!^{\mathrm m}042$.
Hence, the correction of Eq. \ref{procedure} should
not cause large changes in the mean and amplitude of the CPS light curves.
The maximum differences were
$\mathrm{max}(|M(\tau)-M'(\tau)|)=0.\!\!^{\mathrm m}014$ and
$\mathrm{max}(|A(\tau)-A'(\tau)|)=0.\!\!^{\mathrm m}027$.
When we applied PSM to the corresponding CPS parameters as
in Sect. \ref{Cycle},
the results were
\begin{center}
\begin{tabular}{ccc}
                      & $P'_{\mathrm{c}}\pm \sigma'_{P_C}$   & $F'$                  \\
$M'(\tau)$            & $13.\!\!^{\mathrm y}1 \pm 0.\!\!^{\mathrm y}2$  & $1 \times 10^{-10}$ \\
$A'(\tau)$            & $12.\!\!^{\mathrm y}0 \pm 0.\!\!^{\mathrm y}3$  & $5 \times 10^{-4}$ \\
$M'(\tau)+A'(\tau)/2$ & $12.\!\!^{\mathrm y}9 \pm 0.\!\!^{\mathrm y}2$  & $2 \times 10^{-10}$ \\
$M'(\tau)-A'(\tau)/2$ & $18.\!\!^{\mathrm y}2 \pm 0.\!\!^{\mathrm y}4$  & $5 \times 10^{-9}$ \\
\end{tabular}
\end{center}
The best cycles
were practically the same as those for the
original $V$ magnitudes.
All this supported the presence of an activity cycle of
approximately $12^{\mathrm{y}}.5$.

\subsection{Differential rotation in $V'$ 
photometry \label{Diffconv}}

\begin{figure}
   \centering
   \includegraphics[width=\hsize]{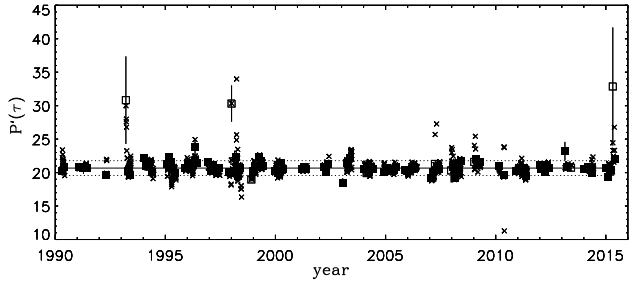}
      \caption{Period $P'(\tau)$ for the corrected $V'$ magnitudes,
otherwise as in Fig. \ref{Pe}
              }
         \label{Peconv}
   \end{figure}

The results for $P'(\tau)$ are shown in Fig. \ref{Peconv}.
The $n'=105$ independent and reliable period estimates had
$P'_{\mathrm{w}}\pm \Delta P'_{\mathrm{w}}=20\fd67 \pm 0\fd37$,
which was equal to $Z'=0.11\equiv 11\%$.
The signal to noise ratio $A/N$ was practically the same as for the
original $V$ magnitude data, because there were no significant
changes in the light curve amplitudes.
Hence, the $V'$ magnitude data gave
close to
the same differential rotation
coefficient estimate $k' \ge Z'_{\mathrm{phys}}=0.08$
(with changes $Z_{\rm spu}=0.08$).
Comparison of Figs. \ref{Pe} and \ref{Peconv} revealed that
the MLC correction eliminated numerous unrealistic
$P(\tau)$ values obtained in the CPS analysis of original $V$ magnitudes.
Most of the remaining unrealistic $P'(\tau)$ values were obtained during
the year 1998 when the amplitude $A'(\tau)$
of light curves was very close to zero.

   \begin{figure}
   \centering
   \includegraphics[width=\hsize]{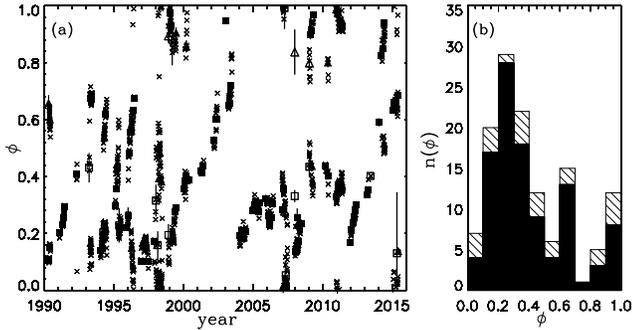}
      \caption{$t'_{\mathrm{min,1}}(\tau)$ and  $t'_{\mathrm{min,2}}(\tau)$ 
for the corrected $V'$ magnitudes
               with active longitude ephemeris of Eq. \ref{AcLo}.
              }
         \label{MiMiconv}
   \end{figure}

   \begin{figure}
   \centering
   \includegraphics[width=\hsize]{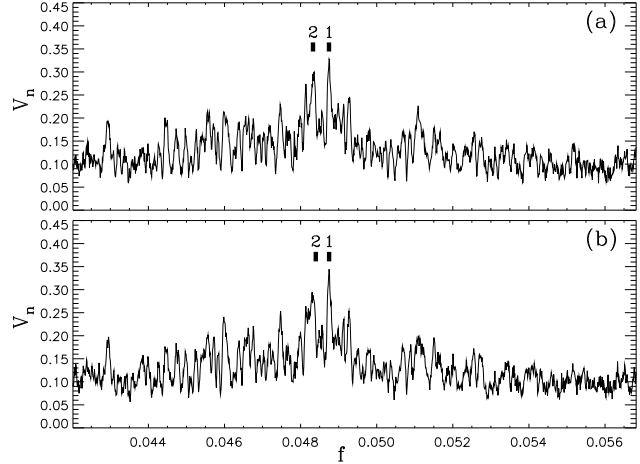}
      \caption{{\bf a}) Kuiper periodogram for the $n=107$ 
independent and reliable
$t_{\mathrm{min,1}}(\tau)$ estimates of original $V$ magnitudes.
The thick horizontal lines denote the locations of the two best periods
$1 \equiv 20\fd51$ and  $2 \equiv 20\fd68$.
                {\bf b}) Kuiper periodogram for the $n'=105$ independent and reliable
$t'_{\mathrm{min,1}}(\tau)$ estimates of corrected $V'$ magnitudes.
The two best periods are $1 \equiv 20\fd51$ and
$2 \equiv 20\fd70$.
              }
         \label{Kuiper}
   \end{figure}

\subsection{Active longitudes in 
$V'$ photometry \label{Activeconv}}

The nonweighted Kuiper test was applied
to the $n'=105$ reliable $t'_{\mathrm{min,1}}(\tau)$ estimates
of independent datasets.
The tested period range was the same as in Sect. \ref{Active}.
The best active longitude rotation period was
$P'_{\mathrm{al}}=20\fd512 \pm 0\fd005$
$(Q_{\mathrm{K}}=5 \times10^{-8})$.
The result for the $n'=129$ reliable
$t'_{\mathrm{min,1}}(\tau)$ and  $t'_{\mathrm{min,2}}(\tau)$
estimates of independent datasets
was the same,
$20\fd512 \pm 0\fd010$ ($Q_{\mathrm{K}}=6 \times 10^{-7}$).
The $P_{\mathrm{al}}$ and $P'_{\mathrm{al}}$ values were also the same
within their error limits.
Therefore, the ephemeris of Eq. \ref{AcLo} was also used
in our Fig. \ref{MiMiconv}, which shows the phases
of $t'_{\mathrm{min,1}}(\tau)$ and  $t'_{\mathrm{min,2}}(\tau)$.
Most of the reliable $t'_{\mathrm{min,1}}(\tau)$ estimates
of independent datasets concentrate
between phases 0.1 and 0.7.

The Kuiper test periodograms $V_{\mathrm{n}}$ for the primary minima of
the original $V$ magnitudes and the corrected $V'$ magnitudes
are shown in Figs. \ref{Kuiper}.
The two best active longitude periods for the original $V$ magnitudes were
$20\fd51$ $(Q_{\mathrm{K}}=1\times10^{-7})$ and
$20\fd68$  $(Q_{\mathrm{K}}=1\times10^{-5})$ .
The $V_{\mathrm{n}}$ peak of the second best period value
in Fig. \ref{Kuiper}a was also very significant.
The two best active longitude periods for the corrected
$V'$ magnitudes were
$20\fd51$ $(Q_{\mathrm{K}}=5\times10^{-8})$ and
$20\fd70$  $(Q_{\mathrm{K}}=4\times10^{-5})$ (Fig. \ref{Kuiper}b).
After the MLC correction, the significance of the best period increased,
while the significance of the second best period decreased.
From this we could conclude that the MLC did not cause the active longitudes.
On the contrary, the MLC hampered the detection of active longitudes.

\section{Conclusions \label{Conc}}

We analysed a quarter of a century of photometry of \object{BM CVn}.
The original $V$ magnitude data were binned as a function of
phase computed with the orbital period ephemeris of Eq. \ref{Elli}.
This revealed that the photometry contained a mean light curve, or MLC,
having a peak to peak amplitude of
 $0.\!\!^{\mathrm m}042$ (Fig. \ref{BinVr}a).
The corrected $V'$ magnitudes were computed by subtracting the MLC
from the original $V$ magnitudes (Eq. \ref{procedure})

The Continuous Period Search, or CPS, algorithm was applied to the original $V$ magnitudes 
and the corrected $V'$ magnitudes.
When we removed the MLC from the data, the fraction of unreliable
CPS models decreased from 8.6\% to 7.4\%.
The same activity cycle of approximately 12.5 years 
was detected with the power spectrum
method from the $V$ and $V'$ magnitudes 
(Sects. \ref{Cycle} and \ref{Cycleconv}).
The results for the differential rotation coefficient
were nearly the same for the $V$ and $V'$ magnitudes,
$k \ge Z_{\mathrm{phys}}=0.06$ and
$k' \ge Z'_{\mathrm{phys}}=0.08$
(Sects. \ref{Diff} and \ref{Diffconv}).
However, the number of unrealistic period estimates
decreased after the MLC was removed (Figs. \ref{Pe} and \ref{Peconv}).
It has to be noted that the estimated level of spurious period changes at $Z_{\rm spu}=0.08$ is significant due to the long rotation period of the star in relation to reasonable dataset lengths. Thus the estimated differential rotation values have considerable uncertainties.
The same active longitude rotation period,
$P_{\mathrm{al}}=20\fd511 \pm 0\fd005$,
was detected with the nonweighted Kuiper method
from the primary minima of $V$ and $V'$ data (Eq. \ref{AcLo}).
This result did not change when the secondary minima were
also included into the analysis (Sects. \ref{Active} and \ref{Activeconv}).
We showed that the MLC hampered the detection of active longitudes.
Hence, the MLC was certainly not the cause for observing such long--term
regularities in the light curve minima of \object{BM CVn}.

The ellipticity of the primary component of another RS CVn star,
$\sigma$ Gem, causes a regular MLC 
as a function of the orbital period \citep{Roe15}.
Due to projection effects, this MLC has two minima and maxima.
\citet{Roe15} argued that this MLC may be the reason for observing
two active longitudes in $\sigma$ Gem, i.e. this phenomenon may not
be connected to dark spots or to chromospheric activity in general.
The MLC of \object{BM CVn} has only one minimum 
and one maximum (Fig. \ref{BinVr})
and ellipticity cannot therefore be the cause for this regularity.
Innumerable different types of light curves have been observed
in different classes of variable stars \citep[e.g.][]{Dra14}.
For example, a mass transfer induced bright spot in
\object{BM CVn} could explain the observed MLC.
If the rate of this mass transfer varies,
the brightness changes caused by the bright spot also vary,
and our MLC correction of Eq. \ref{procedure} cannot remove
such irregularity from the original $V$ magnitude data.
{Nevertheless, we stress that the removal of the 
MLC increased the detectability
of the active longitudes. This is an indication of the MLC being caused by
physics unrelated to spot activity.}
%









\begin{acknowledgements}
This research at the Department of Physics (University of
Helsinki)  was  performed  in  collaboration  with  the  participants
of  the  course “Variable  stars”,  which was lectured  in autumn 2014.
We have made use of the SIMBAD database at CDS, Strasbourg,
France and NASA's Astrophysics Data System (ADS) bibliographic services.
The automated astronomy program at Tennessee State University
 has been supported by NASA, NSF, TSU and the State of Tennessee
through the Centers of Excellence program.
\end{acknowledgements}

\bibliographystyle{an}
\bibliography{siltalareferences}
\end{document}